\documentclass[apjl]{emulateapj}

\usepackage{amssymb}
\usepackage{natbib}
\usepackage{graphicx}

\usepackage{subfigure}
\def\aap{A\&A }
\def\mnras{MNRAS }
\def\apj{ApJ }
\def\pasa{Publ. Astron. Soc. Aust.}

\def\na{New Astronomy}

\shorttitle{The Mysterious Sickle Object in the Carina Nebula}
\shortauthors{Ngoumou et al.}

\begin{document}

\title{The Mysterious Sickle Object in the Carina Nebula:  \\
A stellar wind induced bow shock grazing a clump?}

\author{Judith Ngoumou\altaffilmark{1}, Thomas Preibisch\altaffilmark{1}, Thorsten Ratzka\altaffilmark{1}, Andreas Burkert \altaffilmark{1}}
\affil{Universit\"ats-Sternwarte M\"unchen, Ludwig-Maximilians-Universit\"at, Scheinerstr.1, 81679 M\"unchen, Germany}
\email{ngoumou@usm.lmu.de}

\begin{abstract}
Optical and near-infrared images of the Carina Nebula
show a peculiar arc-shaped feature, which we call the "Sickle",
 next to the B-type star \textit{Trumpler~14 MJ~218}.
We use multi-wavelength observations to explore and constrain the nature
and origin of the nebulosity.
Using sub-mm data from APEX/LABOCA as well as \textit{Herschel} far-infrared maps,
we discovered a dense, compact clump with a mass of
$\sim 40\,{\rm M}_\sun$ located close to the apex of the Sickle.
We investigate how the B-star MJ~218, the Sickle, and the clump are related.
Our numerical simulations show that, in principle,
a B-type star located near the edge of a clump can produce a crescent-shaped
wind shock front, similar to the observed morphology.
However, the observed proper motion of MJ~218 suggest that the star moves
with high velocity ($\sim 100\,{\rm km\,s^{-1}}$)
 through the ambient interstellar gas.
We argue that the star is just about to graze along the surface
of the clump, and the Sickle is a bow shock induced by the
stellar wind, as the object moves supersonically through the density gradient
in the envelope of the clump.
\end{abstract}


\section{Introduction}
The Carina Nebula is a complex star forming region situated at a distance of $2.3$~kpc from the sun.  With more than $70$ massive stars identified \citep{2007MNRAS.379.1279S}, it is an ideal site to study the impact of stellar feedback on the interstellar medium (ISM). In this highly dynamical environment massive stars affect the surrounding medium through photoionizing radiation, stellar winds and ultimately through supernova explosions. Observations and numerical simulations have shown that these mechanisms can produce various observable morphologies like superbubbles at large scales \citep{2004ApJ...613..302O, 2011ApJ...731...13N},  cavities \citep{2012ASPC..453...25F}, shells \citep{2010A&A...523A...6D, 2012MNRAS.427..625W} or pillars \citep{2010ApJ...723..971G, CNC-LABOCA, 2012MNRAS.427..625W}.  The feedback mechanisms at work during the lifetime of the massive stars (before the supernova) may not play the most dominant role on larger scales \citep{2011MNRAS.414..321D}. However, these processes can affect the star formation locally by either triggering the formation of new stars \citep{2010ApJ...723..971G, 2012MNRAS.427..625W, 2012A&A...540A..81O} or dispersing clouds and thereby delaying or even hindering star formation \citep{2012MNRAS.427..625W}.  

In addition to the feedback mechanisms mentioned above, the dynamical evolution of these massive stars is also likely to affect the surroundings. While most of the massive stars are believed to lie within their native OB association \citep{2003ARA&A..41...57L} and are likely to be part of a binary or multiple system \citep{2007ApJ...670..747K, 2001IAUS..200...69P, 1998AJ....115..821M}, some stars with quite high velocities, so called runaways, have been observed outside of OB associations \citep{1961BAN....15..265B, 2012AJ....143...71K, 2012MNRAS.424.3037G}.
 
In this article we take a look at multi-wavelength observations of a peculiar nebulosity around a star in the Carina Nebula and the identification of a denser molecular clump at the same location. We interpret the crescent-shaped nebulosity,  which we call the 'Sickle',  as the tip of a bow shock associated with the B1.5 V star \textit{Trumpler~14 MJ~218} (MJ~218 hereafter) listed in \cite{1993AJ....105..980M}. We discuss a possible link between the star, the Sickle and the clump and argue that the star is moving supersonically through the ambient density gradient on the front side of the observed compact clump.

The object caught our attention because of its peculiar morphology in our inspection 
of optical images of the Carina Nebula region. 
It is located about $1'$ south-east of the dense young cluster Tr~14,
which corresponds to a projected physical distance of about 0.8~pc.

A literature search showed that the peculiar nebulosity 
had already been mentioned in the near-infrared (NIR) imaging study 
of the Tr~14 region by \cite{2007A&A...476..199A}, who suggested the idea of a compact HII region 
around the star. 
The highly asymmetric shape of the nebula could be the result of the
irradiation from the very luminous early
O-type  stars in the center of Tr~14 (most notably the O2If star HD 93129A),
since the apex of the crescent points toward this direction.
\cite{2010MNRAS.406..952S}, however, noted that 
the lack of detectable H$\alpha$ emission is in conflict
with the interpretation as an HII region; they rather argued that the
crescent nebula is a dusty bow shock.

Our investigation of the comprehensive multi-wavelength data set of the Carina Nebula
that we compiled during the last years shows that the crescent nebula
also appears to be related to a compact clump, which is 
(naturally) invisible in the optical and NIR images, but quite prominent
in our far-infrared and sub-mm data. This makes the nebula even more interesting.

\section{Observational data and results}
\begin{figure*}[htp]
  \centering
  \subfigure{\includegraphics[scale=0.5]{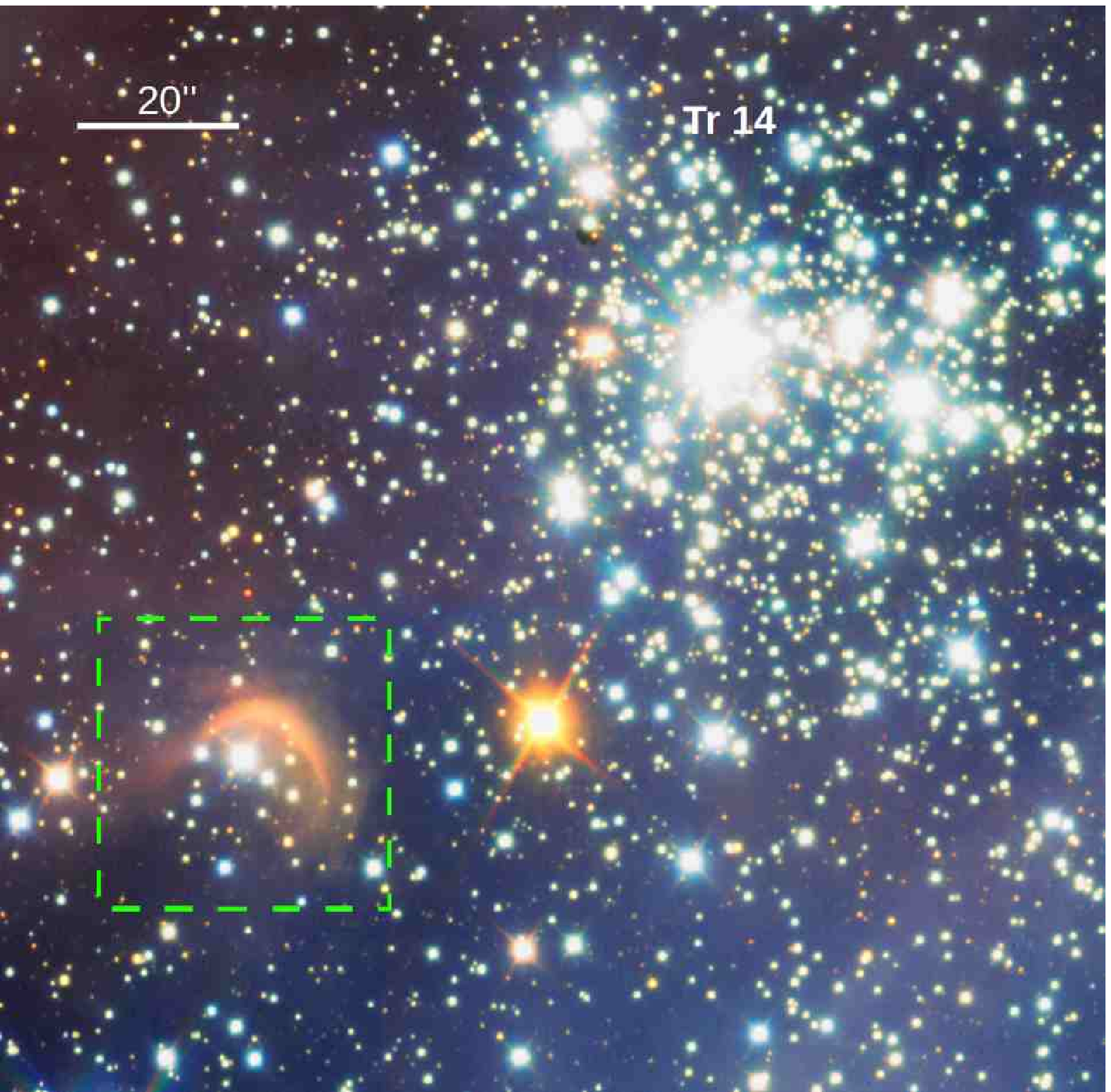}}\qquad
  \subfigure{\includegraphics[scale=0.5]{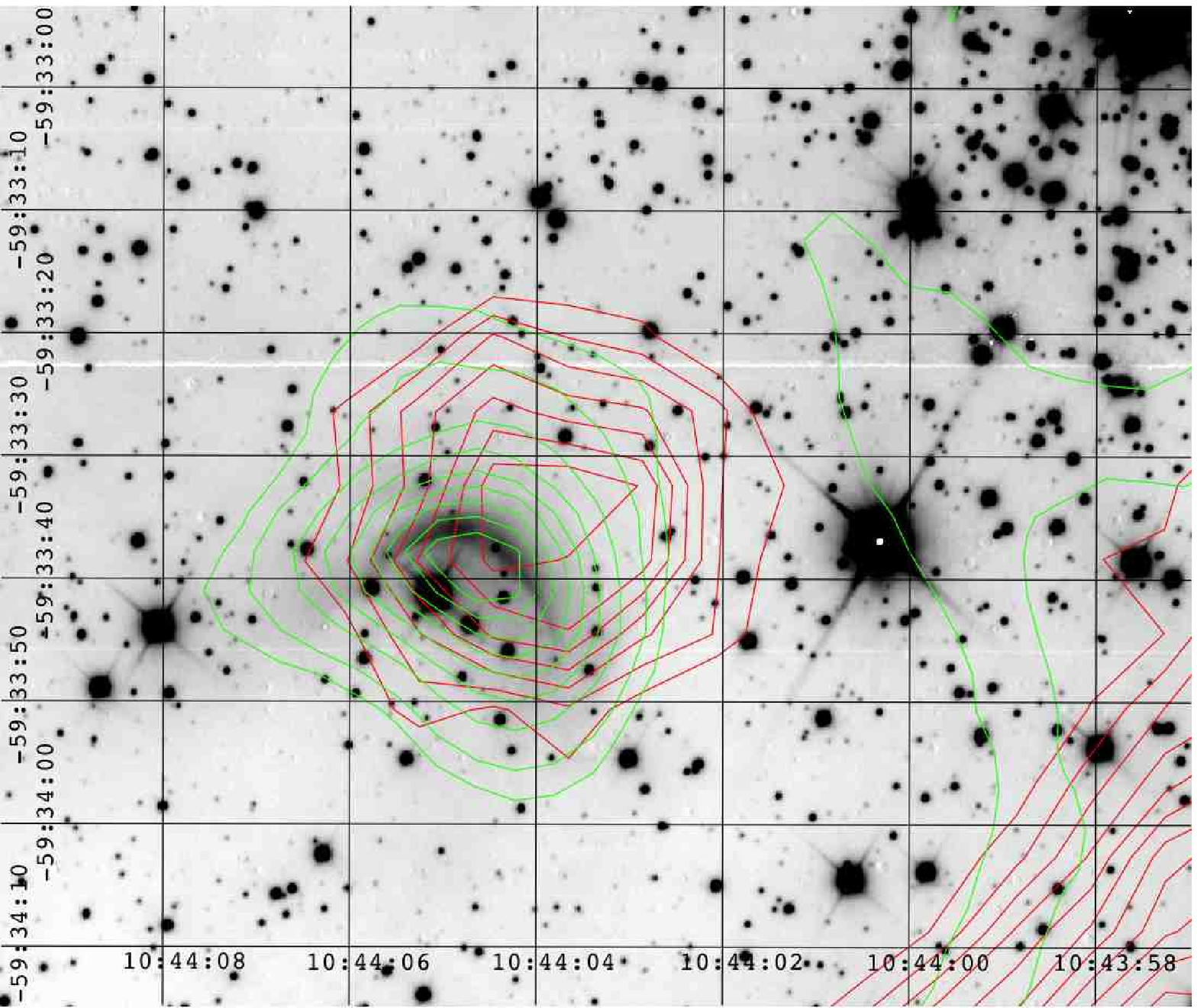}}
\caption{left: RGB composite image constructed from the $J$- (blue),
$H$- (green), and $K_s$-band (red) HAWK-I images of the area around the
Sickle object (marked by the green dashed box in the lower left part)
 and the cluster Tr~14
(see also ESO photo release 1208;
http://www.eso.org/public/news/eso1208/).\\
right:
 Negative grayscale representation of the
$K_s$-band HAWK-I image of the area around the Sickle object
with superposed contours of our \textit{Herschel}
$70\,\mu$m map (green) and our APEX/LABOCA $870\,\mu$m map (red).
The contours levels of  \textit{Herschel} $70\,\mu$m  map
are at 2.25, 2.5, 3.0, 3.5, 4.0, 4.5, 5.5, and 6.0 Jy/pixel
(pixel size $3.2''$); the rms noise level in the map is $\sim 2.3$~Jy/pixel.
The contour levels of the APEX/LABOCA $870\,\mu$m
map go from 0.05 Jy/beam to 0.5 Jy/beam in steps of 0.05 Jy/beam;
as the rms noise level in the map is $\sim 0.02$~Jy/beam, the first
contour corresponds to the $\sim 2.5\,\sigma$ level.
Note that the horizontal bright streak in the background image
is an artifact related to the dither pattern and the mosaicing process.}
 \label{fig1}
\end{figure*}

\paragraph{Stellar parameters}

The observed crescent is associated to the B1.5 V star
 MJ~218 as listed by \cite{1993AJ....105..980M} in their spectroscopic and photometric analysis of the stars in and around the clusters Trumpler~14 and Trumpler~16. MJ~218 alias  
 \textit{2MASS~J10440508-5933412}, alias \textit{ ALS~19740} in \cite{2003AJ....125.2531R} is
located at the J2000 coordinates $10^h\,44^m\,05.1^s$,  $-59\degr\,33'\,41''$, at about $1'$ south-east from the center of Tr~14.
With optical/NIR magnitudes of $V=11.85$ and $K=9.63$ the star is
a bright and prominent object.
The stellar spectral type was determined via optical spectroscopy by \cite{1993AJ....105..980M}. 
These properties are very well consistent with the assumption that this star is 
a member of the Carina Nebula at a distance of 2.3~kpc.

In the UCAC4 Catalogue \citep{2012yCat.1322....0Z, 2013AJ....145...44Z} \footnote{see
http://www.usno.navy.mil/USNO/astrometry/
optical-IR-prod/ucac}, the star
is listed as UCAC4~153-055048. Its proper motion is given as
${\rm pm(RA)} = - 7.0  \pm 3.0\, {\rm mas}\,\, {\rm yr}^{-1}$ and
${\rm pm(Dec)} = 5.2  \pm 3.7\, {\rm mas}\,\, {\rm yr}^{-1}$.
The total proper motion of $8.7 \pm 4.8\, {\rm mas}\,\, {\rm yr}^{-1}$
corresponds to $95 \pm 52\,{\rm km\, s}^{-1}$. This is a remarkably high
velocity, but we have to note that the uncertainties are quite large.
Using the radial velocity of $v_{\rm rad} =  -10.9\,{\rm km}\,{\rm s}^{-1}$ \citep{2006ApJ...648..580H}
leads to a total space velocity of $v_{\rm *} \approx  96\,{\rm km}\,{\rm s}^{-1}$.
This large velocity suggest that MJ~218 is a runaway star.

We note that the amplitude and direction of the motion would be consistent
with the idea that the star MJ~218 could have been ejected some 66\,000 years ago from the region of the
open cluster Tr~16 which is at a distance of  $\sim 6.5$~pc.

\begin{figure*}[htp]
\centering
\includegraphics[scale= 0.95]{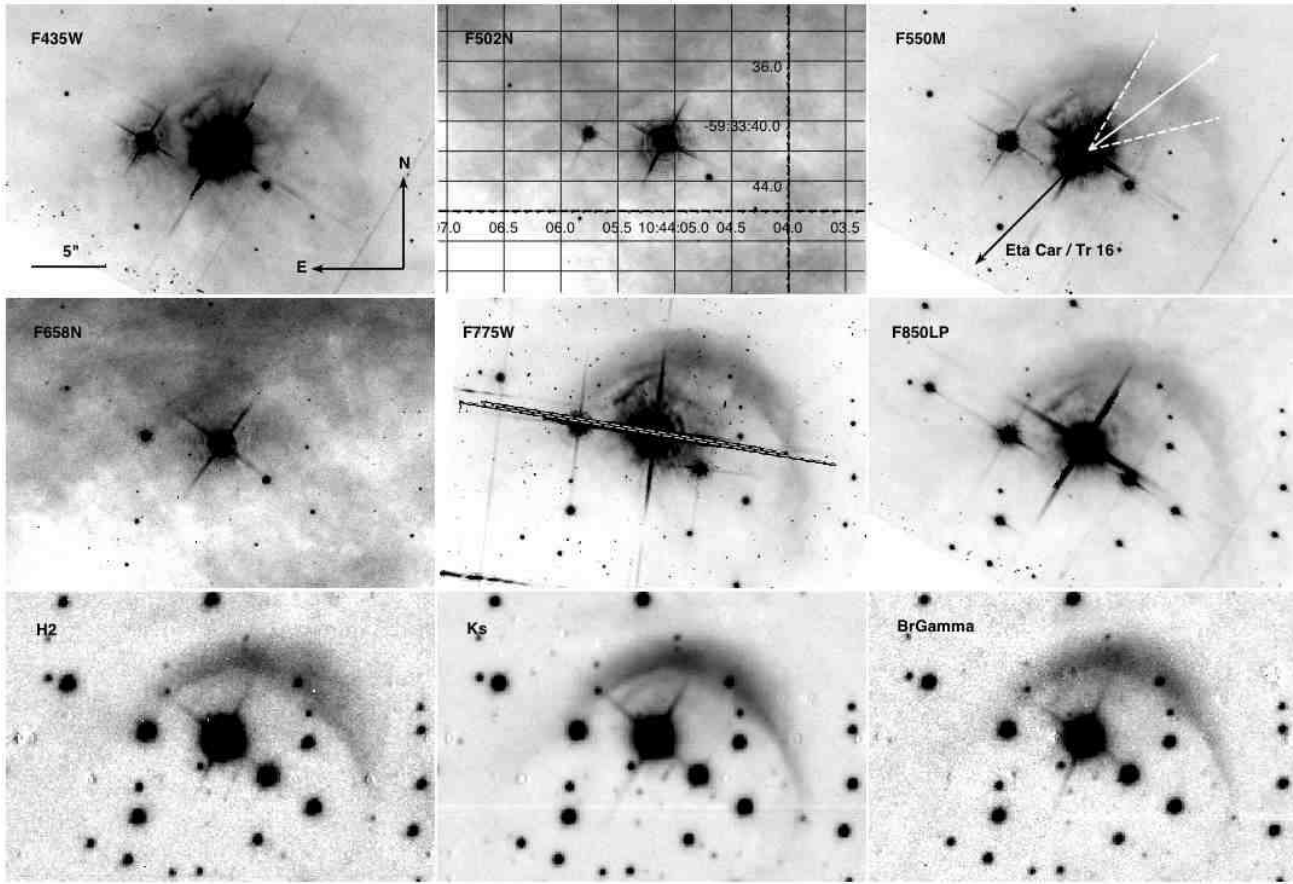}
 \caption{Images of the object taken with the HST (top and middle) and HAWK-I (bottom). A linear scale is used. The direction of the catalogued proper motion is indicated as a white arrow in the third panel (upper right). The uncertainty in the position angle of the velocity vector is indicated by the white dashed lines.  The black arrow denotes the direction to $\eta$ Carinae in Tr~16.
In the fifth panel the prominent double diffraction spike is marked, indicating a bright compact source close to the B1.5V star.\label{HSTImg}.
  }
\end{figure*}

\paragraph{\textit{Hubble Space Telescope} optical images}

We searched the \textit{Hubble} Legacy Archive for observations of the Sickle object. The images shown in Fig.~\ref{HSTImg} are taken with the Advanced Camera for Surveys (ACS). The dates of the observations, the exposure times, and the proposal identifiers are given in Table~\ref{HSTObs}. Fig.~\ref{HSTImg} shows the compilation of HST
in different filters, which reveal the small-scale structure of the
nebulosity in detail. Two additional shells are visible 
closer to the star. Whereas the nebulosity is very well visible in the
broad-band filters,
neither the F502N band filter (tracing the [O~III] line) nor the
F658N filter (tracing the H$\alpha$ line) reveal significant diffuse emission,
clearly showing that the emission from the crescent nebula is not line-emission
but continuum emission.

\begin{table}
\caption{Journal of HST observations.}
\label{HSTObs}
\centering
\resizebox{\columnwidth}{!}{
\begin{tabular}{cccl}
\hline\hline\noalign{\smallskip}
\hline\noalign{\smallskip}
Filter & Date       & Exp. Time[s] & Proposal IDs\\
\noalign{\smallskip}\hline \noalign{\smallskip}
F435W  & 2006-07-29 &   698	   &  10602 (PI: J.Maiz Appellaniz)\\
F502N  & 2010-02-01 &  7650	   &  12050 (PI: Mario Livio)\\
F550M  & 2006-07-29 &   678	   &  10602 (PI: J.Maiz Appellaniz)\\
F658N  & 2005-07-17 &  1000	   &  10241 (PI: Nathan Smith)\\
F775W  & 2003-05-17 &	550	   &   9575 (PI: William B. Sparks)\\
F850LP & 2006-07-29 &   678	   &  10602 (PI: J.Maiz Appellaniz)\\
\noalign{\smallskip}\hline
\end{tabular}
}
\end{table}

\paragraph{Near-infrared images}

To investigate the near-infrared morphology of the nebula,
we inspected data obtained
in January 2008 as part of our survey of the Carina Nebula
\citep[see][for more details]{CNC-HAWKI}
with the instrument HAWK-I at the
ESO 8m Very Large Telescope.
Images were obtained in the
standard $J$-, $H$-, and $K_s$-band filters, as well as
in narrow-band filters centered on the $2.121\,\mu$m
  $\nu=1\!-\!0$\, S(1) ro-vibrational emission line of
molecular hydrogen
and the $2.166\,\mu$m Bracket $\gamma$ line.
The very good seeing conditions during these observations 
resulted in sub-arcsecond FWHM values of typically
$0.37''$ (in the $K_s$-band) for the PSF size of
point-like sources near the Sickle object.

Comparison of the narrow-band images with the $K_s$-band image
shows:
In the Bracket $\gamma$ line, there is no increase of the
brightness of the nebula relative to the star MJ~218. This excludes significant hydrogen line emission.
Furthermore, there is also no indication of Bracket $\gamma$ 
rim-brightening at the northern edge of the Sickle. Such a rim-brightening
is clearly visible in narrow-band images in many of the irradiated globules in the Carina Nebula  as well as in optical and near infrared  images of photoevaporating globules and pillars \citep{2003ApJ...587L.105S,2004MNRAS.351.1457S}. Rim-brightening  highlights the strong surface irradiation by the
surrounding hot stars. For the Sickle nebula, however,
external irradiation seems not to be important.

In the molecular hydrogen line narrow-band image the
relative brightness of the Sickle nebula is not much larger than
in the broad-band image, but the nebulosity seems to be slightly
more extended at the north-western edge. This suggests moderate amounts
of molecular hydrogen line emission at the front of the nebula,
similar as often observed in the bow-shocks of protostellar jets
that move through molecular clouds \citep{CNC-HAWKI, 2011MNRAS.416.2163T}.

\paragraph{Mid-Infrared data: MSX and TIMMI2}

The Sickle nebula coincides with the Midcourse Space Experiment (MSX)
 point source MSX6C~G287.4288-00.5804.
\cite{2007A&A...476.1019M} performed mid-infrared imaging
of this source with better angular resolution at the 3.6~m ESO
Telescope and resolved the MSX source
into four MIR sources. Their $10.4\,\mu$m image clearly shows
the Sickle nebula (but no sign of the star MJ~218), and the peak of the
emission corresponds very well to the brightest point of the 
 Sickle as seen in our $K_s$-band image.
 \cite{2007A&A...461...11U}  and \cite{2007A&A...476.1019M}  reported a non-detection in radio continuum emission from this source. \cite{2007A&A...461...11U} stated a detection limit  of  $0.4 $~mJy. 
Assuming  an electron temperature of  $10^4$~K and the distance of $2.3$~kpc, we derived an excitation parameter of $ \sim 2\, {\rm pc}\,{\rm cm}^{-2}$. By extrapolating Table 14.1 in \cite{2009tra..book.....W}, this value appears roughly consistent with a B1.5 star.

\paragraph{LABOCA sub-mm map}

Whereas the optical and NIR images show no indication for the
presence of a dense clump in the surroundings of the Sickle nebula,
our $870\,\mu$m map that was obtained in December 2007
with the bolometer array LABOCA at the APEX telescope \citep[see][for a complete
description of this data set]{CNC-LABOCA} clearly reveals a compact
clump located close to the tip of the Sickle.
The position of the peak of the sub-mm emission is 
 $10^h\,44^m\,04.06^s$,  $-59\degr\,33'\,35.5''$, i.e.~10 arcseconds (or 0.1~pc)
north-west of the star MJ~218
and about 3 arcseconds north-west of the apex point of the
Sickle nebula. The sub-mm emission is almost point-like
and only marginally extended; given the $18''$ angular resolution
of LABOCA, the width of the clump is 
 $\sim 0.2$~pc.
The peak of the clump emission  has an intensity of 0.393 Jy/beam.

\paragraph{\textit{Herschel} far-infrared maps}

Maps of the Carina Nebula at the wavelengths of
70, 160, 250, 350, and 500~$\mu$m were obtained in
December 2010 in the Open Time project OT1-tpreibis-1
using the parallel
fast scan mode at $60''/{\rm s}$ for simultaneous imaging
with PACS \citep{2010A&A...518L...2P}
and SPIRE \citep{2010A&A...518L...3G}.
 A full description of these observations and the
subsequent data processing can be found in \cite{CNC-Herschel}.
The angular resolution of the \textit{Herschel} maps is
$5''$, $12''$, $18''$, $25''$,
and $36''$ for the 70, 160, 250, 350, and $500\,\mu$m
band, respectively. At a distance of 2.3 kpc this corresponds to
physical scales ranging from 0.06 to 0.4~pc.

The clump near the Sickle nebula detected by LABOCA 
is clearly visible in the 
\textit{Herschel} maps. It is clearly extended in the PACS maps,
where we determine FWHM values of $19'' \times 16''$.
These maps also show that the shape of the clump is not
central symmetric. It exhibits a kidney-shaped form (green contours in Fig.~\ref{fig1}, right panel), with a slight caved-in side in south-east direction in accordance with the inner side of the
Sickle nebula bow. This clearly suggests that the process
creating the Sickle nebula bow interacts with the clump.

Using the methods described in \cite{CNC-Herschel}, we determined
the column density, temperature, and mass of the clump.
The peak value for the column density in the center
of the clump is $N_{\rm H} \approx 1.3 \times 10^{22}\,\rm cm^{-2}$,
corresponding to a visual extinction of $A_V \approx 6.5$~mag 
assuming a normal extinction law (note that these numbers are beam-averaged values, i.e.~the
true values for a line-of-sight exactly through
the center will be higher). The dust temperature in the
clump is found to be about 32~K.
The mass of the clump can be determined by integrating the
column density over all pixels exceeding the limit 
$A_V > 3$~mag (in order to separate the clump from the surrounding 
diffuse gas); this yields $M_{\rm clump} \approx 40\,M_\sun$.

\paragraph{X-ray data}

The object is located in the area covered by the
{\it Chandra} Carina Complex Project, that recently
mapped a 1.3 square-degree region of the Carina Nebula.
With an exposure time of $\sim 60$~ksec
($\sim 17$ hours) for the individual mosaic positions,
the on-axis completeness limit is $L_{\rm X} \approx 10^{29.9}$~erg/s
in the $0.5-8$~keV band for lightly absorbed sources.
A complete overview of the {\it Chandra} Carina Complex Project
 can be found in
\citet{CCCP}, which is the introduction to
a set of 16 papers resulting from this project.

In the \textit{Chandra} images analyzed in the context of the CCCP,
the star MJ~218
is clearly detected as an X-ray point-source with 59 source counts.
The J2000 position of the X-ray source is $10^h\,44^m\,05.09^s$,  $-59\degr\,33'\,41.4''$ and
has a total 1-$\sigma$ error (individual source position error and systematic astrometrical uncertainty)
circle radius of $\approx 0.4''$ \citep{2011ApJS..194....2B}.
This position agrees perfectly (i.e.~within less than $0.1''$)
with the optical position of the star listed
in the UCAC4 catalog ($10^h\,44^m\,05.091^s$,  $-59\degr\,33'\,41.37''$).
The positional offset to the corresponding 2MASS counterpart is $0.2''$, i.e.~well
within the uncertainties.
From our inspection of the optical HST image we found an (insignificant) offset of
$0.1''$ between the star and the X-ray source position; the nearest other star
visible in the  HST image is $3.8''$ offset from MJ~218.
In the near-infrared HAWK-I images, we found an (insignificant) offset of $0.2''$,
and a distance of $1.9''$ to the nearest other point-source. We therefore conclude that we
have a clear and unambiguous
identification of the X-ray source with the star MJ~218.

The X-ray properties of this source can be summarized as follows
\citep[see][for details]{2011ApJS..194....2B}:
the median photon energy of the source is 1.48~keV.
The analysis of the photon arrival times yields some, although rather weak
evidence for variability: the Kolmogorow-Smirnow test gives a probability
of $P_0 = 0.16$ for the null hypothesis of a constant count rate.
The fit to the X-ray spectrum with XSPEC yielded as plasma
temperature of $kT = 2.5 (\pm 0.8)$~keV and gave an extinction-corrected
intrinsic X-ray luminosity of  $L_{\rm X} \approx 8.3 \times 10^{30}\,\rm erg/sec$.

According to the well established results for the origin of stellar X-ray emission
\citep[see, e.g.][]{2009A&ARv..17..309G}.
no X-ray emission is expected for MJ~218,
since stars of spectral type B1.5 should neither show coronal magnetic activity
as typical for late-type (spectral types F and later) stars, nor should they
have sufficiently strong stellar winds, in which X-ray emission is produced in
wind shocks, as observed in the (much more  luminous) O-type stars.
 The general lack of intrinsic X-ray emission from stars in the spectral type
range from $\sim$~B1 to late A has been well confirmed in numerous
X-ray observations \citep [e.g.,][] {1985ApJ...290..307S, 2002ApJ...578..486D, 2005ApJS..160..401P,
2005ApJS..160..557S}.

The common explanation for the detections of X-ray emission from  B stars
is thus the assumption that the emission actually originates from an unresolved
late-type companion \citep[e.g.,][]{2011ApJS..194...13E}.
The observed median photon energy and the plasma temperature of MJ~218
derived from the X-ray spectrum are fully consistent with this
assumption, and considerably higher than one would expect from wind-shock related
X-ray emission \citep[e.g.,][]{2000ARA&A..38..613K}.
Considering the general correlation between X-ray luminosity and stellar mass
for young stars \citep[see][]{2005ApJS..160..401P}, the observed X-ray luminosity
suggests the companion to have a stellar mass around
$\sim 1-2\,M_\odot$.
Given the above described upper limit for a possible angular offset of the
X-ray source from the B-star position of $\la 0.2''$, the putative low-mass companion
must have a projected separation of less than $\sim 460$~AU from the B-star.
This rather small value makes a chance projection highly unlikely.\footnote{In order to
quantify this statement, we inspected the HST image and counted the number of detectable
stars within $10''$ of MJ~218 to be 8. With this estimate of the local star density,
the Poisson probability to find one or more unrelated stars as chance projection
within $0.2''$ of MJ~218 is just 0.3\%.} Since many B-type stars are known to
have lower-mass companions at separations of a few ten to a few hundred AUs
\citep[see, e.g.,][]{1999NewA....4..531P, 2007A&A...474...77K, 2013A&A...550A..82G}
the hypothesis of a binary system seems to be the best explanation of the observed parameters.\\
We searched in the HST images for such a companion, but even the narrow-band images are saturated at the required small distances from the position of MJ 218 due to the brightness of the star.\footnote{The HST images (Fig.~\ref{HSTImg})  show a double diffraction spike indicative of a companion separated by about 0.2 arcsec in north-south direction. The similar brightness of this potential companion, however, disqualifies it as X-ray source.}

\section{Theoretical considerations}

\subsection{Clump carving scenario} \label{incloud}

\begin{figure}[h]
\resizebox{ \columnwidth}{!}{\includegraphics{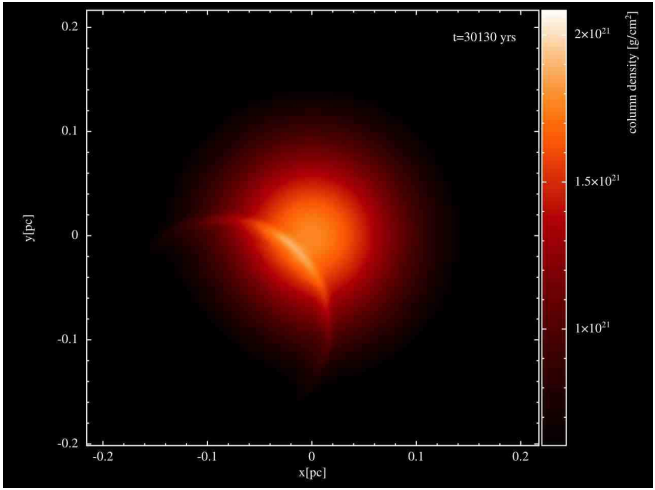}}
\caption {Integrated density plot  for a projected star-clump distance $d_{obs}=0.11$~pc  at an inclination of  $\alpha=66\degr$.  The image was produced using SPLASH \citep{2007PASA...24..159P} 
\label{clumpcarv}}
\end{figure}

The evolution of the wind bubble around a star located at the edge of a clump can lead to a crescent-like shaped shock front. 
We performed a simulation using the Smoothed Particle Hydrodynamics (SPH) code SEREN \citep{2011A&A...529A..27H} including a newly implemented HEALPix-based \citep{2005ApJ...622..759G} momentum conserving stellar wind-scheme (Ngoumou et al. 2013, in preparation) to simulate the expansion of a momentum driven wind bubble in a molecular clump.   The choice of momentum driven \citep{1975ApJ...198..575S} as opposed to thermal pressure driven \citep{ 1975ApJ...200L.107C, 1977ApJ...218..377W} is justified, as the stellar winds of a spectral type B1.5 are assumed to be too weak to induce a hot shocked, X-ray emitting layer. 

The clump was modeled as a supercritical Bonnor-Ebert sphere \citep{1956MNRAS.116..351B, 1957ZA.....42..263E}, which is a self-gravitating isothermal sphere confined by an external pressure. The sphere has the dimensionless radius $\xi = 8$ and a finite radius $R_\mathrm{BES}\approx 0.31$~pc, corresponding to a FWHM size of $\sim 0.2$~pc and a central density of  $\rho_\mathrm{BES}\approx 1.7\times10^{-19}\,{\rm g}\,{\rm  cm^{-3}}$ at a temperature of $36$~K, for a total mass of $M_{\rm clump}= 40 \, {\rm M}_\sun$. A static momentum source was placed at the edge of the nebula at a distance of  $0.29$~pc from the center of the clump.  
We used  the mass loss rate and terminal wind velocity for a B1.5 V star as stated by \cite{2006MNRAS.367..763S}:  $\dot{M}_{\rm w}= 6\times10^{-8}\, {\rm M}_\sun\,{\rm yr}^{-1}$ and $v_{\infty} = 960\,{\rm km}\,{\rm s}^{-1}$.  

After a few $10^4$~yrs, the crescent has reached the size of the observed object.  
To match the observed projected distance $d_{obs}\approx 0.11$~pc between the star and the center of the clump (inferred from the sub-mm map), the simulation box is rotated by an angle $\alpha=66\degr$  around the axis passing the center of the clump and perpendicular to the clump-star axis.  Fig.~\ref{clumpcarv} shows the density integrated along the line of site at $t= 3\times10^4$~yrs when the crescent has reached the size of the Sickle. The inclination angle between the plane containing the star and the center of the clump, and the projection plane is $\alpha= 66\degr$. 

 This scenario though, requires the star to be embedded inside the clump. It seems unlikely that MJ~218 was born inside or close to the clump as it would certainly have been dispersed by the stellar wind by now. The assumption that the star is coming from somewhere else and travelled through the ISM, finding itself embedded inside the clump would require wind shell radii smaller than the radius of the clump.  Shells with larger  radii would overrun the clump, and compress it \citep{2011ApJ...736..142B,2010ApJ...723..971G,2012A&A...546A..33T} leading to a cometary/pillar like structure with the tail pointing away from the star.  This is not observed.
 The required compact shell could be produced naturally if the star would move supersonically through the ISM. In this case,
 a star produces a bow shock with smaller radii at the collision front of the wind with the ambient medium, which we discuss in the next section.
  
\subsection{Bow schock scenario} \label{bowschock}

\begin{figure}[h]
\raggedright
\resizebox{ \columnwidth}{!}{\includegraphics[angle=0]{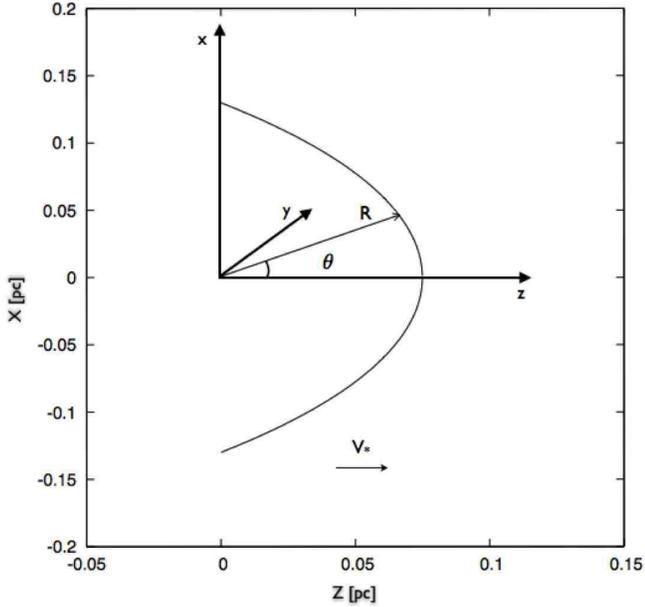}}
\caption{Schematic diagram of a stellar wind bow shock to illustrate the definition of the coordinate system.  \label{coord}}
\end{figure}

The reported velocity of MJ~218 is very high ($\sim 96\,{\rm km}\,{\rm s}^{-1}$). The errors, however, are of the order of $55\%$, making the value rather uncertain.  A star with such high velocity forms a bow shock while traveling through an ambient medium with temperatures $T \le 10^6$~K.  Indeed, the position of the tip of the arc approximately correlates with the direction of the velocity vector in the plane of the sky as inferred from proper motion measurements of the star (see Fig.~\ref{HSTImg}). To test this scenario we compare the distance between the star and the cusp of the Sickle nebula with the stand-off radius $R_0$ inferred from the analytical solution for the shape of a stellar wind bow shock in the thin-shell limit as derived in \cite{1996ApJ...459L..31W}. $R_0$ depends on the velocity of the star $v_\mathrm{*}$, on the wind mass loss rate $\dot{M}_\mathrm{w}$, on the terminal wind velocity $v_\mathrm{w}$ and on the ambient density $\rho_\mathrm{AMB}$.   
\begin{equation}
R_0 = \sqrt{\frac{\dot{M}_\mathrm{w} v_\mathrm{w}}{4\pi\rho_\mathrm{AMB}v^2_\mathrm{*}}} 
\label{eq1}
\end{equation}
The shape of the bow shock near the stand-off radius is given by:
\begin{equation}
R_{\theta} = R_0\, {\rm cosec } \, \theta \sqrt{3\left(1-\theta \cot \theta \right)} 
\label{eq2}
\end{equation}
with $\theta$ being the polar angle measured from axis given by the direction of motion of the star (see Fig.~\ref{coord}).

\begin{figure}[h]
\raggedright
\resizebox{ \columnwidth}{!}{\includegraphics[angle=0]{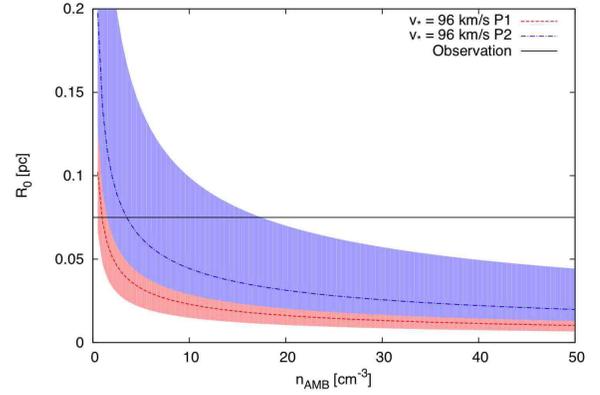}}
\caption{Stand-off radius of the bow shock (Eq. \ref{eq1}) against number density of the ambient medium for $v_\mathrm{*} = 96\,{\rm km}\,{\rm s}^{-1} $ and for two sets of stellar wind parameters; P1 (red dot-dashed): $\dot{M}_{\rm w} = 1.1\times10^{-8}\, {\rm M}_\sun\,{\rm yr}^{-1}$ and $v_{\infty} = 1400\,{\rm km}\,{\rm s}^{-1}$; P2 (blue dashed): $\dot{M}_{\rm w}= 6\times10^{-8}\, {\rm M}_\sun\,{\rm yr}^{-1}$ and $v_{\infty} = 960\,{\rm km}\,{\rm s}^{-1}$. The shaded area represent the spread due to the $55\%$ error on  $v_\mathrm{*}$. 
The horizontal solid line indicates the stand-off radius inferred from observations.
 \label{standoff}}
\end{figure}

Fig.~\ref{standoff} shows the variation of the stand-off radius of the bow shock with ambient density  (Eq.~\ref{eq1}) for a range of stellar velocities and for two sets of wind parameters  P1(red dot-dashed) and P2(blue dashed).
\begin{description}
\item [{\rm P1}] $\dot{M}_{\rm w} = 1.1\times10^{-8}\, {\rm M}_\sun\,{\rm yr}^{-1}$;  $v_{\infty} = 1400\,{\rm km}\,{\rm s}^{-1}$ \citep[based on models computed by][]{2001A&A...375..161P}  \footnote{see~ 
 http://www.usm.uni-muenchen.de/people/adi/Models/
 Model.html}
\item [{\rm P2}]  $\dot{M}_{\rm w}= 6\times10^{-8}\, {\rm M}_\sun\,{\rm yr}^{-1}$;  $v_{\infty} = 960\,{\rm km}\,{\rm s}^{-1}$ \citep{2006MNRAS.367..763S}
\end{description}
The shaded area represent the spread due to the $55\%$ error on  $v_\mathrm{*}$. 
The horizontal solid line indicates the stand-off radius inferred from the reported proper motion and radial velocity measurements which indicate an inclination angle of $\sim 6 \degr$ and a value of  $R_0$ of $\sim 0.075$~pc. The star is almost moving in the plane of the sky.

The observed stand-off distance is obtained for number densities ranging between $\sim 0.1\,{\rm cm}^{-3}$ and  $\sim 20\,{\rm cm}^{-3}$. The range of ambient densities is given by the large errors on the velocity estimate for MJ~218 (shaded area in Fig.~\ref{standoff}). 
For the reported $\sim 96\,{\rm km}\,{\rm s}^{-1}$, $n_{AMB}= 2\,{\rm cm}^{-3}$ for P1 and $n_{AMB}= 8\,{\rm cm}^{-3}$ for P2. These values are consistent with an order-of-magnitude estimate for the density of the rather diffuse gas
in the inner parts of the Carina Nebula superbubble, through which the star is moving. This hints at the Sickle being the bow shock induced by  MJ~218 moving supersonically through the diffuse ISM and not directly interacting with the densest part of the clump.

The question now arises whether we just see a star with a bow shock projected in front of an unaffected clump. 
Interestingly, the bow shock does not appear to be symmetric around the axis given by the velocity vector of MJ~218 (see upper right panel in Fig.~\ref{HSTImg}) as expected if the star would move in isolation. We therefore suggest that we indeed see a contribution from the clump. 
 \cite{2000ApJ...532..400W} investigated the modifications of bow shocks of stars running into an ambient density gradient in the direction perpendicular to the stellar motion and the effect of anisotropic winds.  In both cases he found configurations in which  the star does not lie on the symmetry axis dividing the bow shock into two parts.  The observed asymmetry for the position of MJ~218 with respect to the tip of the Sickle might therefore be an indication of an interaction of the moving star with the clump.  
\begin{figure}[h]
\resizebox{ 1\columnwidth}{!}{\includegraphics[angle=0]{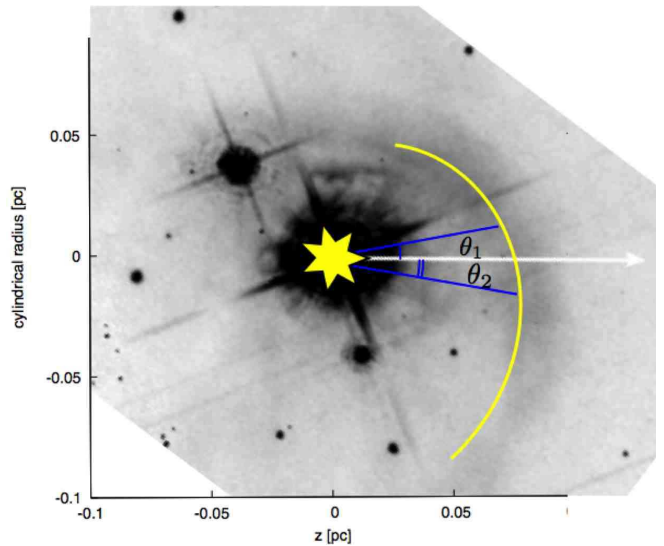}}
\caption{Illustration of a  bow shock solution in a linear stratified medium  with slope $a\approx15\,{\rm pc}^{-1}$ for $R_0 = 0.075$~pc overlaid on an HST image (F550M band filter; image is rotated by $36\degr$). The velocity vector points in $z$-direction.
\label{asybow}}
\end{figure}
As an illustration we consider the solution for a bow shock of a star moving in  a linear stratified medium with a density gradient perpendicular to the stellar velocity vector as described in \cite{2000ApJ...532..400W}:
 \begin{equation}
\rho_{AMB} = \rho_0\left(1+ay\right)
\label{eq3}
\end{equation}
The characteristics of the solution near the stand-off point for a bow shock seen from the side (i.e.~ $v_\mathrm{*}$ perpendicular to the line-of-sight) are given by (see Eq. (71) in \cite{2000ApJ...532..400W}):
 \begin{equation}
\frac{R_{\theta}}{R_0} =  1 - \frac{aR_0}{4}\theta + \left[\frac{1}{5} + \frac{5}{32}a^2 R_0^2 \right]\theta^2 
\label{eq4}
\end{equation} 
We approximate the gradient  $a$ by selecting two points on the observed bow and measure their respective angles  $\theta_1$ and  $\theta_2$  and the corresponding distances  $R_{\theta_1}$ and  $R_{\theta_2}$ from the star. From Eq.~\ref{eq4} we estimate to first order:
 \begin{equation}
a=\frac{4}{R_0^2} \left(\frac{R_{\theta_1}-R_{\theta_2}}{\theta_2-\theta_1}\right)
\label{eq5}
\end{equation}
 Fig.~\ref{asybow} shows the result of this approximation (yellow arc) for $R_{\theta_1} \approx 0.95 R_0$ at  $\theta_1 \approx 10\degr$  and for $R_{\theta_2} \approx 1.05 R_0$ at  $\theta_2 \approx -10\degr$.  The model illustrates the non axis-symmetric shape of the bow shock  induced by the additional term in the linear term near the stand-off point.

The asymmetric shape of the Sickle could be the result  of MJ~218 running in a medium stratified in the direction perpendicular to its motion which coincides with the line-of-sight and the position in the sky of the clump. The observed clump could therefore be part of the stratification and be located behind the Sickle.

\section{Conclusion} 

The Sickle nebulosity in the Carina Nebula is by itself already a very
remarkable and interesting feature.
The discovery of a dense cloud clump at a projected location
just in front of the Sickle prompted us to investigate a possible relation
 between these two features.
  We simulated the impact of the momentum transfer from the stellar wind  on a clump with the star being located at the edge of the clump. 
The off-center expansion of the wind bubble could create a crescent-like front which resembles the observed Sickle nebula. This scenario, however, requires a static source embedded in the clump.  The star would have to be formed  in the last $10\,000$~yrs at this position. This seems very unlikely. We therefore exclude this scenario.
 
Measurements of the proper motion of MJ~218 taken from the UCAC 4  catalogue \citep{2012yCat.1322....0Z} indicate a high velocity of nearly $v_\mathrm{*} = 96\,{\rm km}\,{\rm s}^{-1} $.  At this velocity the wind of MJ~218 would form a bow shock while traveling through an ambient medium with temperatures $T \le 10^6$~K.  The Sickle could be part of the bow shock 
of MJ~218 seen from the side, as MJ~218 appears to be moving almost in the plane of the sky. Assuming the stand-off radius to be the distance between the star and the intersection of its velocity vector with the Sickle, we measure a stand-off radius $R_0\approx 0.075$~pc. Such values for $R_0$ point at ambient densities in the range of  $\sim 0.1\,{\rm cm}^{-3}$ and  $\sim 20\,{\rm cm}^{-3}$ which are far lower than the $10^4\,{\rm cm}^{-3}$ inferred for the center of the observed clump. 
The star is thus not moving through the inner, central parts of the clump, but seems to be grazing along the surface of the clump.

The non-axisymmetric appearance of the Sickle is consistent with
the presence of a density gradient perpendicular to the direction of
motion of MJ~218, which coincides with the line-of-sight and the location
 of the clump.  We therefore argue that the observed Sickle is part of the
bow shock of the high velocity B-star MJ~218 grazing the front surface
density gradient of the observed compact clump. The asymmetry of the
Sickle with respect to MJ~218 can then be explained as
 a result of this interaction.

We expect the Sickle to be a rather transient feature which is likely to evolve on timescales of order  $10^4$~yrs. 
Surely more detailed observation of this peculiar object will help to better constrain its nature.  Spectral information from the molecular material could probe the velocity structure inside the clump and help test our star-clump interaction scenario. 
An interesting question is also whether the passage of the star could trigger the gravitational collapse of the clump and lead to star formation. This issue will be addressed in a subsequent paper.
In addition a closer look at the binary nature of MJ~218 and its implications could help shed a light on the complex history of the Carina Nebula. Especially  the question of the origin MJ~218 in the framework of the formation of massive runaway binaries \citep{2007ApJ...660..740M, 2011A&A...529A..14G,2012MNRAS.424.3037G} would be worth investigating.   
The direction of the velocity vector of MJ~218
suggests a possible origin in the open cluster Trumpler~16 in the center
of the Carina Nebula. The observed proper motion would then imply a
travel time of about $10^5$~yrs.
It is very interesting to note that the back-projected motion path
puts MJ~218 in close vicinity of a recently detected
neutron star candidate in the Carina Nebula,
described in \cite{2009ApJ...695L...4H}.
This raises the possibility that MJ~218 has perhaps been ejected in a
relatively recent supernova explosion in the Carina Nebula.
Further investigations of this relations could  thus provide interesting
information on the dynamical evolution, stellar clustering
\citep{2010MNRAS.404..721M} and the still very unclear history of
past supernovae in the Carina Nebula.

\begin{acknowledgements}
This publication makes use of data obtained with the \textit{Herschel} spacecraft.
The \textit{Herschel} spacecraft was designed, built, tested, and launched under a
contract to ESA managed by the \textit{Herschel}/Planck Project team by an industrial
consortium under the overall responsibility of the prime contractor Thales
Alenia Space (Cannes), and including Astrium (Friedrichshafen) responsible for
the payload module and for system testing at spacecraft level, Thales Alenia
Space (Turin) responsible for the service module, and Astrium (Toulouse)
responsible for the telescope, with in excess of a hundred subcontractors."

This publication is partly based on data acquired with the Atacama Pathfinder Experiment (APEX). APEX is a collaboration between the Max-Planck-Institut fur Radioastronomie, the European Southern Observatory, and the Onsala Space Observatory.

Based on observations made with the NASA/ESA \textit{Hubble} Space Telescope, and
obtained from the \textit{Hubble} Legacy Archive, which is a collaboration between the
Space Telescope Science Institute (STScI/NASA), the Space Telescope European
Coordinating Facility (ST-ECF/ESA) and the Canadian Astronomy Data Centre
(CADC/NRC/CSA).

This project is funded by the German
 \emph{Deut\-sche For\-schungs\-ge\-mein\-schaft, DFG\/} PR 569/9-1. Additional support came from funds from the Munich Cluster of Excellence: "Origin and Structure of the Universe".
 
We would like to thank  Jim Dale for the very helpful comments.
\end{acknowledgements}

\end{document}